\documentclass[preprint,showpacs,preprintnumbers,amsmath,amssymb]{revtex4}

\usepackage{graphicx}

\begin{document}

\title{
Possible daily and seasonal variations in quantum interference 
induced by Chern-Simons gravity 
}
\author{Hiroki Okawara}
%\email{}
\author{Kei Yamada}
%\email{}
\author{Hideki Asada} 
%\email{asada@phys.hirosaki-u.ac.jp}
\affiliation{
Faculty of Science and Technology, Hirosaki University,
Hirosaki 036-8561, Japan} 

\date{\today}

\begin{abstract}
Possible effects of Chern-Simons (CS) gravity 
on a quantum interferometer turn out to be 
dependent on the latitude and direction 
of the interferometer on the Earth 
in orbital motion around the Sun. 
Daily and seasonal variations in phase shifts are predicted 
with an estimate of the size of the effects, 
wherefore neutron interferometry 
with $\sim 5$ meters arm length and 
$\sim 10^{-4}$ phase measurement accuracy  
would place a bound on 
a CS parameter 
comparable to Gravity Probe B satellite. 
\end{abstract}


\maketitle

%\section{Introduction}
\noindent \emph{Introduction.--- } 
It has long been a fundamental issue 
to understand the interplay between the quantum theory and 
the gravitational physics. 
The interplay is studied mostly by theoretical experiments 
\cite{Refs}. 
Corella, Overhauser, and Werner (COW) \cite{COW}
succeeded an first experiment involving 
both the Plank constant $h$ and the gravitational constant $G$ 
by using a neutron interferometer.  
In COW experiments, a neutron interferometer is tilted, 
such that a neutron beam path I is higher above the surface of
the Earth than the other path segment II, 
causing a gravitationally induced phase shift of the neutron 
de Broglie waves on path II relative to path I. 
The gravitationally induced phase shift was experimentally 
observed \cite{COW, Sakurai-Book}. 
In recent years, technological progress has been brought into 
quantum experiments including neutron interferometers 
and quantum optics. 
Current attempts to probe general relativistic effects 
in quantum mechanics focus on 
precision measurements of phase shifts in 
quantum interferometers (e.g. \cite{Zych}). 
Hogan has recently proposed an ambitious idea 
to use quantum interferometers as an experimental probe 
of a quantum spacetime at the Planck scale \cite{Hogan}. 
Quantum experiments may play a role in probing 
an intermediate regime between general relativistic gravity 
and Planck scale physics. 

Current astronomical observations, such as the apparent 
accelerated expansion of the Universe, suggest 
a possible infrared modification to general relativity (GR).
The Chern-Simons (CS) correction is not an {\it ad hoc} extension, 
but it is actually motivated by both string theory, as a
necessary anomaly-canceling term to conserve unitarity
\cite{Polchinski}, 
and loop quantum gravity \cite{Ashtekar}. 
Alexander and Yunes have recently pointed out that 
CS gravity possesses the same parameterized post-Newton (PPN) 
parameters as general relativity, 
except for the inclusion of a new term, proportional to the CS coupling 
and the curl of the PPN vector potential \cite{AY1, AY2}. 
They have also shown 
that this new correction 
might be used in {\it classical} experiments, 
such as Gravity Probe B, to bound CS gravity and test string theory 
(See \cite{AY2009} for an extensive review of CS modified gravity).

In contrast to approaches focusing on general relativistic effects 
on quantum systems \cite{Refs}, we shall study CS gravity 
in quantum experiments as another attempt to probe quantum gravity. 
Nandi and his collaborators \cite{Nandi} 
have recently discussed the quantum phase shift 
in Chern-Simons modified gravity, 
where an isolated gravitating body was considered. 
They have concluded 
that the induced shifts by the spin of the body 
are too tiny to be observed. 
However, the Earth's orbital angular momentum 
($\sim 3 \times 10^{40} \, \mbox{kg} \cdot \mbox{m}^2 \mbox{s}^{-1}$) 
is much larger than its spin angular momentum 
($\sim 7 \times 10^{33} \, \mbox{kg} \cdot \mbox{m}^2 \mbox{s}^{-1}$). 
Both of the axial vectors may play a role in CS gravity. 
Therefore, we consider gravitationally interacting bodies 
in order to investigate the quantum mechanical effects of 
the Earth's orbital angular momentum in CS gravity. 
The main result of this Letter 
suggests that a CS modified gravity theory 
may predict daily and seasonal phase shifts in quantum interferometers, 
which are in principle distinct from the general relativistic effects. 
This feature can be currently used as a {\it quantum} tool 
to probe CS gravity. 

%\section{CS gravity}
\noindent \emph{CS gravity.--- } 
CS gravity modifies GR via the addition of a correction to 
the Einstein-Hilbert action, namely \cite{Jackiw, Guarrera}
\begin{eqnarray}
S_{CS}=\frac{1}{16\pi G}\int d^4x\frac{1}{4}fR^{\star}R , 
\label{CS-action}
\end{eqnarray}
where 
$f$ is a prescribed external field (with units of area in 
geometrized units) that acts as a coupling constant, 
$R$ is the Ricci scalar, and the star stands for the dual operation. 

The weak-field solution to the CS modified field equations 
in PPN gauge is given by \cite{AY1, AY2, AY2009}
\begin{eqnarray}
g_{00}&=& -1 + 2U -2 U^2 + 4 \Phi_1 + 4 \Phi_2 + 2 \Phi_3 
+ 6 \Phi_4 + O(6) , 
\\
g_{0i}&=& -\frac72 V_i - \frac12 W_i +2 \dot f (\nabla\times V)_i 
+O(5), 
\label{g0i}
\\
g_{ij}&=& (1 + 2U) \delta_{ij} + O(4) , 
\end{eqnarray}
where ${U, \Phi_1, \Phi_2, \Phi_3, \Phi_4, V_i, W_i}$ 
are PPN potentials (e.g. \cite{Will}), 
$O(A)$ stands for PN remainders of order $O(1/c^A)$ 
for the light speed $c$ 
and the dot denotes the derivative with respect to $x^0 \equiv ct$. 
Note that this is a non-dynamical (kinematical) model of 
CS modified gravity, where we assume that $f$ depends on time only. 
The non-dynamical CS theory is tractable 
and could become a good approximation in weak fields, 
regardless of a possible evolution problem of the external field $f$ 
(presumably near the central region) consistent with 
Pontryagin constraint. 
A full dynamical study of seeking approximate solutions 
for rotating extended bodies has yet to be carried out 
\cite{AY2009}. 
Henceforth, we investigate whether $\dot f$ term 
in Eq. (\ref{g0i}) brings new gravitational physics 
into quantum systems. 

Following \cite{AY1}, 
let us consider a system of nearly spherical bodies 
in the standard PPN point-particle approximation. 
A follow-up study conducted by Smith and his collaborators \cite{Smith} 
shows that the new term in Eq. (\ref{g0i}) is valid even outside of 
a weakly gravitating spinning body like the Earth. 
For the above vector potential $V_i$, 
the CS correction to the metric becomes 
in the barycenter frame 
\cite{AY1, AY2, AYcorr, AY3}
\begin{eqnarray}
\delta_{CS} g_{0i}=
\frac{2G}{c^3} \sum_A \frac{\dot f}{r_A} 
\left[\frac{m_A}{r_A}(\vec v_A \times \vec n_A)^i -\frac{J^i_A}{2r^2_A} 
+\frac{3}{2}\frac{(\vec J_A \cdot \vec n_A)}{r^2_A}n^i_A \right] ,
\label{deltag0i}
\end{eqnarray}
with $m_A$ the mass of the Ath body, 
$r_A$ the field point distance to the Ath body, 
$n_A^i = x_A^i/r_A$ a unit vector pointing to the Ath body, 
$v_A^i$ the velocity of the Ath body, 
$J_A^i$ the spin-angular momentum of the Ath body,  
and the $\cdot$ and $\times$ operators are the flat-space 
inner and outer products. 
Note that the CS correction couples  
with the spin 
and the orbital angular momenta. 
%Henceforth, we investigate how this new term 
%affects quantum interferometers. 

\begin{figure}[t]
\includegraphics[width=12cm]{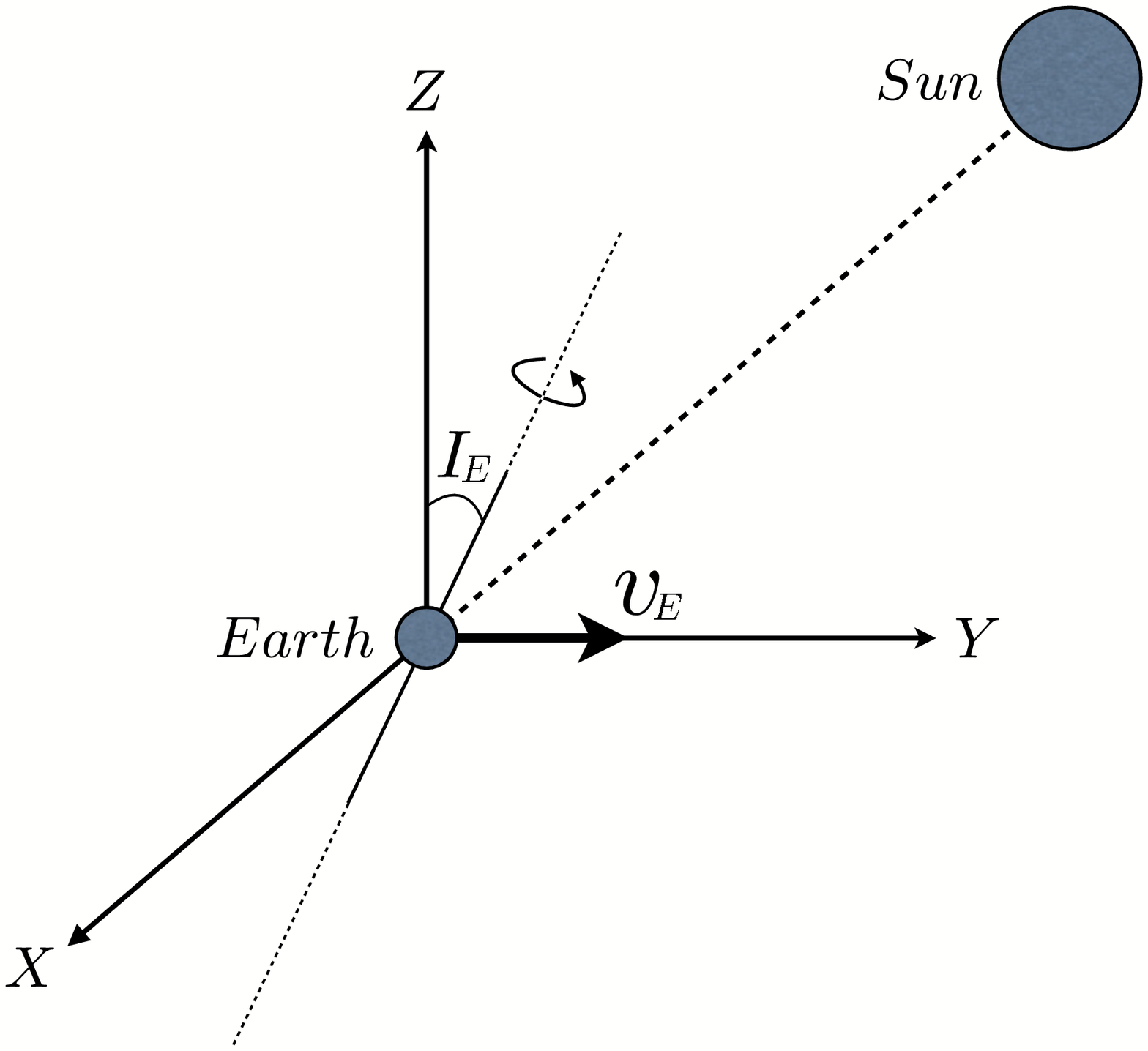}
\includegraphics[width=12cm]{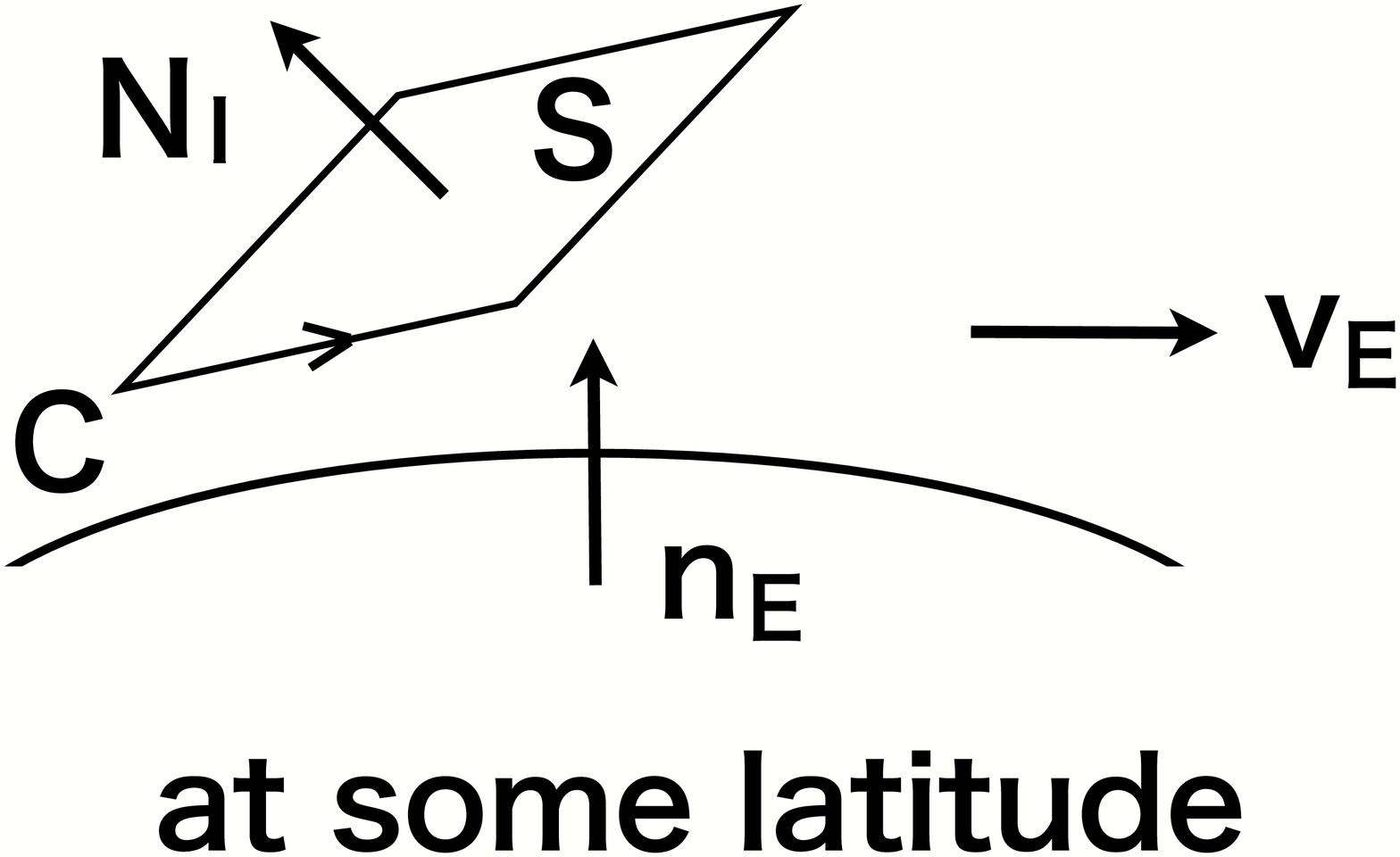}
\caption{ 
Quantum interferometer on the Earth orbiting around the Sun. 
The orbital plane is chosen as the $X-Y$ plane. 
The Earth's axis and orbital velocity are 
denoted by $I_E$ and $v_E$, respectively. 
Top panel: Earth orbiting around the Sun. 
Bottom panel: Interferometer  
at a certain time and place 
on the Earth. 
The latitude and the longitude are specified by $\vec n_E$, 
which rotates in an inertial frame 
around the Earth's axis 
and hence its direction changes also with 
the orbital motion of the Earth. 
The interferometer's direction  
$\vec N_I$ also 
changes in an inertial frame as the Earth rotates. 
}
\label{f1}
\end{figure}

%\section{Phase shifts in a quantum interferometer}
\noindent \emph{Phase shifts.--- } 
We consider a quantum interferometer 
that consists of a closed path $C$ (its area $S$) 
on the Earth, 
as shown by Fig \ref{f1}. 

The Hamiltonian for a quantum particle in a curved spacetime 
involves $g_{\mu\nu}$. 
The linear-order correction to the Hamiltonian by $g_{0i}$ becomes 
$\delta H = mc g_{0i} v^i$ 
for a slowly-moving particle \cite{LL}. 
A phase difference induced by $g_{0i}$ is thus 
expressed as \cite{Sakurai-Book}
\begin{eqnarray}
\Delta&=&
\frac{1}{\hbar} \oint_C \delta H dt 
\nonumber\\ 
&=& 
\frac{mc}{\hbar} \oint_C \vec g \cdot d\vec r , 
\label{Delta1}
\end{eqnarray}
where $\vec g$ denotes $(g_{01}, g_{02}, g_{03})$, 
$m$ denotes the quantum particle mass, 
$\hbar \equiv h/2\pi$ denotes Dirac's constant. 
By using Stokes theorem, $\Delta$ is rewritten in the surface 
integral form over $S$ as 
\begin{eqnarray}
\Delta=\frac{mc}{\hbar}\int_S (\vec \nabla \times \vec g) 
\cdot d\vec S . 
\label{Delta2}
\end{eqnarray}
This form has an analogy in the Aharonov-Bohm (AB) effect. 
The AB effect in the phase shift, 
which was confirmed experimentally 
\cite{Chambers}, is  
$\propto \oint_C \vec A \cdot d\vec r = 
\int_S (\vec\nabla \times \vec A)\cdot d\vec S$ 
for a vector potential $\vec A$ in the electromagnetism.  
Note that 
the phase difference $\Delta$ in Eq. (\ref{Delta2}) 
is caused by time dilation 
and hence it does not depend on de Broglie wavelength $\lambda$, 
in contrast to COW experiments. 

Let us substitute the CS term  
of Eq. (\ref{deltag0i}) 
into Eq. (\ref{Delta2}) to obtain $\Delta$ for CS gravity. 
By using an identity $\epsilon^{ijk} (1/r)_{, jkl} = 0$ 
with the Levi-Civita symbol $\epsilon^{ijk}$, 
one can see that 
the $J$-dependent part of the metric in Eq. (\ref{deltag0i}) 
always vanishes in Eq. (\ref{Delta2}), 
whereas the $v$-dependent part makes contributions. 

Since $\Delta$ involves the curl operation 
on the surface of the Earth 
and the Earth radius $r_E$ is much shorter than 1AU, 
the terms associated with 
the solar mass $M_{\odot}$ in Eq. (\ref{Delta2}) 
is $O(M_{\odot} M_E^{-1} r_E^3 1\mbox{AU}^{-3}) \sim 10^{-9}$ 
smaller than those with the Earth's mass $M_E$, 
so that the terms  
with 
the solar mass (and other planetary ones) 
can be safely neglected. 
Henceforth, we focus on the Earth mass 
(also its spin and orbital angular momentum) 
in CS gravity. 
Hence, Eq. (\ref{Delta2}) becomes  
\begin{eqnarray}
\Delta_{CS}&=&
\frac{2m}{\hbar c^2}\int_S \dot f\frac{GM_E}{r^3} 
\left[3(\vec v_E \cdot \vec n_E)\vec n_E -\vec v_E\right] 
\cdot\vec N_I dS 
\nonumber\\
&=&2\dot f\frac{mGM_ES}{\hbar c^2 r_E^3} 
\left[3(\vec v_E \cdot \vec n_E)\vec n_E -\vec v_E \right] 
\cdot\vec N_I , 
\label{Delta-CS}
\end{eqnarray}
where we used $r_E \gg \sqrt{S} $ 
(the Earth radius is much larger than 
the interferometer arm length) 
and hence $r = r_E$ in the integrand. 
Here, in an inertial frame, 
$\vec v_E$ denotes the Earth's orbital velocity, 
$\vec n_E$ stands for the unit vertical vector on the ground 
(at a certain latitude), 
$\vec N_I$ means the unit normal to the interferometer plane 
(See also Fig. \ref{f1}). 
The unit normal vectors $\vec n_E$ and $\vec N_I$ 
in an inertial frame change with time 
as the Earth rotates. 
The change rate depends on the latitude. 
Moreover, $\vec N_I$ depends also 
on the interferometer's 
direction such as horizontal and vertical. 
In contrast to COW experiments, 
the interferometer direction 
such as North and East does matter in CS gravity. 
Therefore, the factor 
$\left[3(\vec v_E \cdot \vec n_E)\vec n_E -\vec v_E \right] 
\cdot\vec N_I$ in Eq. (\ref{Delta-CS}), 
depending on the latitude and direction, 
changes with the Earth's spin and orbital motion. 

In order to see more explicitly the interplay between quantum mechanics 
and CS gravity, the magnitude of Eq. (\ref{Delta-CS}) is factored as 
\begin{equation}
|\Delta_{CS}| \sim 4 \left(\frac{mc^2}{\hbar}\right) 
\left(
\frac{\dot f}{c} 
\frac{GM_E}{c^2r_E}
\frac{v_E}{c}\right) 
\left(\frac{S}{r_E^2}\right) , 
\label{Delta-CS2}
\end{equation}
where $[3(\vec v_E \cdot \vec n_E)\vec n_E -\vec v_E] \cdot\vec N_I 
\sim 2 v_E$. 
It is worthwhile to mention that the first fraction  
in the right hand side of Eq. (\ref{Delta-CS2}) is 
due to the quantum mechanical physics 
and it is large enough $\sim 10^{24} \, \mbox{s}^{-1}$ 
to compensate the factor in the second parenthesis 
due to the CS gravitational effect 
$\sim \dot f c^{-1} \times 10^{-14}$, 
where $m$ is neutron mass. 
The last factor in Eq. (\ref{Delta-CS2}) 
is the squared ratio of the interferometer 
arm length (often $\sim 60 \, \mbox{cm}$) 
to the Earth radius. 
In total, the magnitude of $\Delta_{CS}$ is 
\begin{equation}
|\Delta_{CS}| \sim 10^{-3} \mbox{s}^{-1} \times 
\left(\frac{mc^2}{1 \mbox{GeV}}\right) 
\left(\frac{\dot f}{c}\right) 
\left(\frac{S}{0.4 \mbox{m}^2}\right) . 
\label{Delta-CS3}
\end{equation}

On the other hand, 
COW experiments that measure phase shifts due to Newton gravity 
rely on the inclination angle 
of the interferometer but not on the latitude 
\cite{COW}. 
Moreover, the general relativistic effects 
of a slowly rotating object, known as 
the Lense-Thirring effects, cause a phase shift 
proportional to $\vec \omega_E \cdot \vec S$ 
(e.g. \cite{Kuroiwa, Wajima}), 
wherefore they depend only on 
the angle between the Earth's axis 
and the interferometer. 
Hence, the directional dependence of the general relativistic 
phase shifts is in principle distinct from 
that of CS gravity.  

\begin{figure}[t]
\includegraphics[width=12cm]{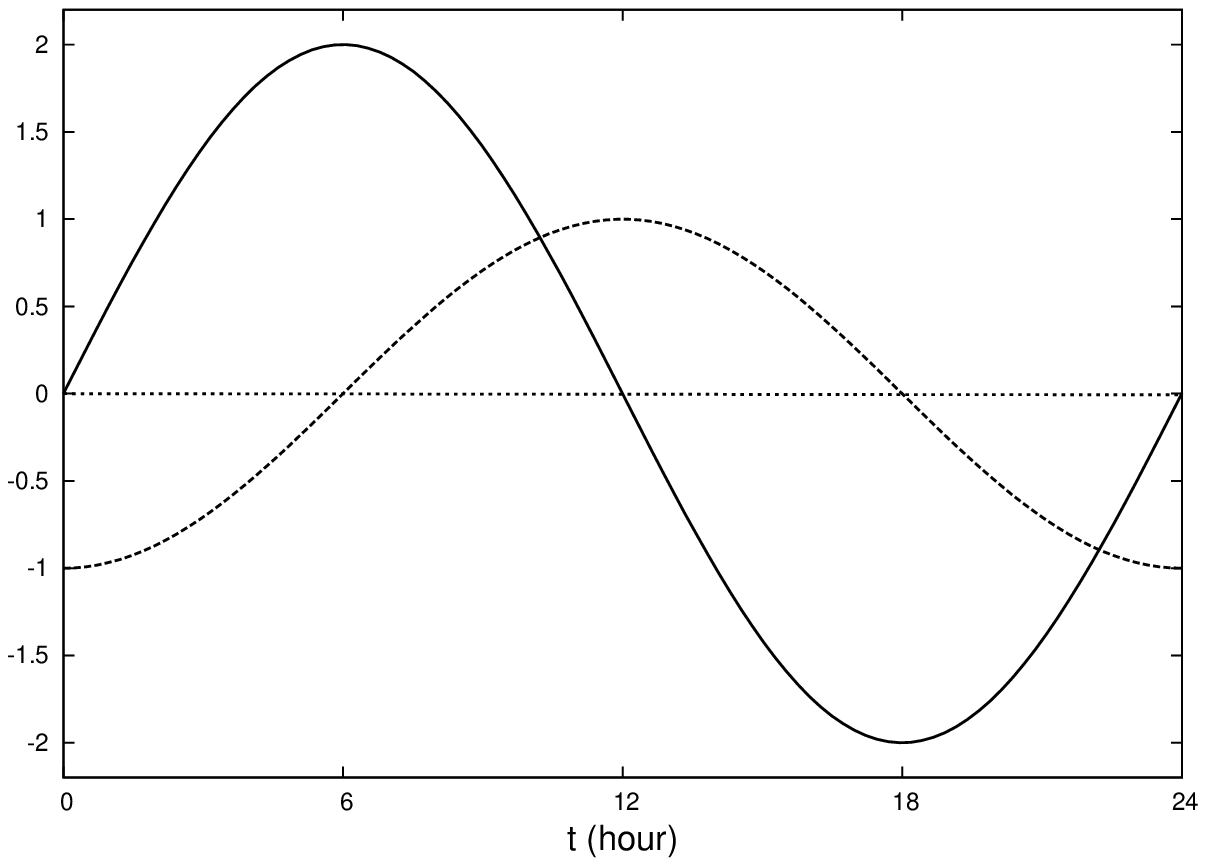}
\caption{ 
Time variation in phase differences by CS effects 
on the vernal equinox day. 
The vertical axis (in arbitrary units) denotes 
$\left[3(\vec v_E \cdot \vec n_E)\vec n_E -\vec v_E \right] 
\cdot\vec N_I$  
in Eq. (\ref{Delta-CS}). 
The quantum interferometer is located 
on the equator of the Earth.  
We consider three cases of the interferometer direction. 
The solid, dashed and dotted curves 
correspond to $\vec N_I$ for a horizontal plane 
and two vertical ones (one facing the East and 
the other facing the North), respectively. 
The midnight is chosen as 0 hour. 
At midnight and at noon 
on the same day, CS effects on 
the phase difference vanish only for the horizontal case. 
This vanishing can be 
shown also by using Eq. (\ref{Delta-CS}), 
because $\vec v_E \perp \vec n_E \parallel \vec N_I$. 
}
\label{f2}
\end{figure}

\begin{figure}[t]
\includegraphics[width=12cm]{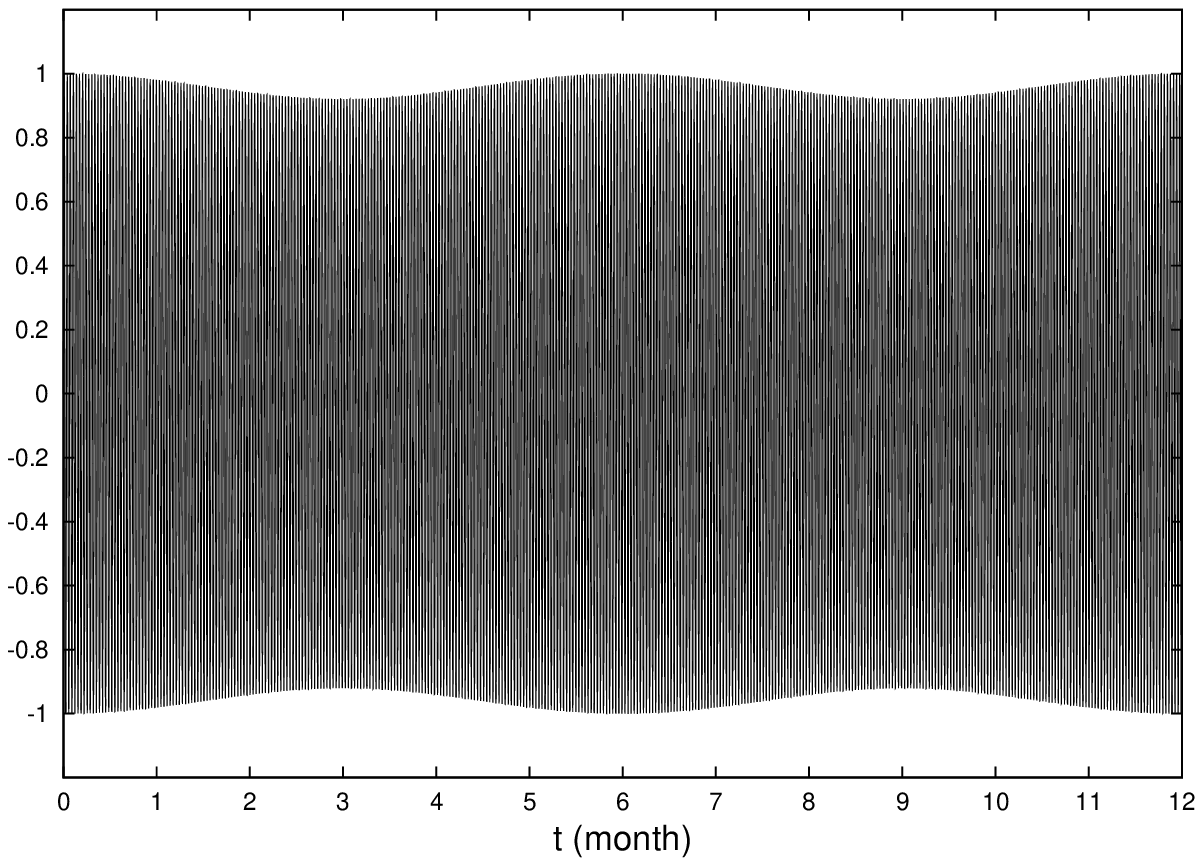}
\includegraphics[width=12cm]{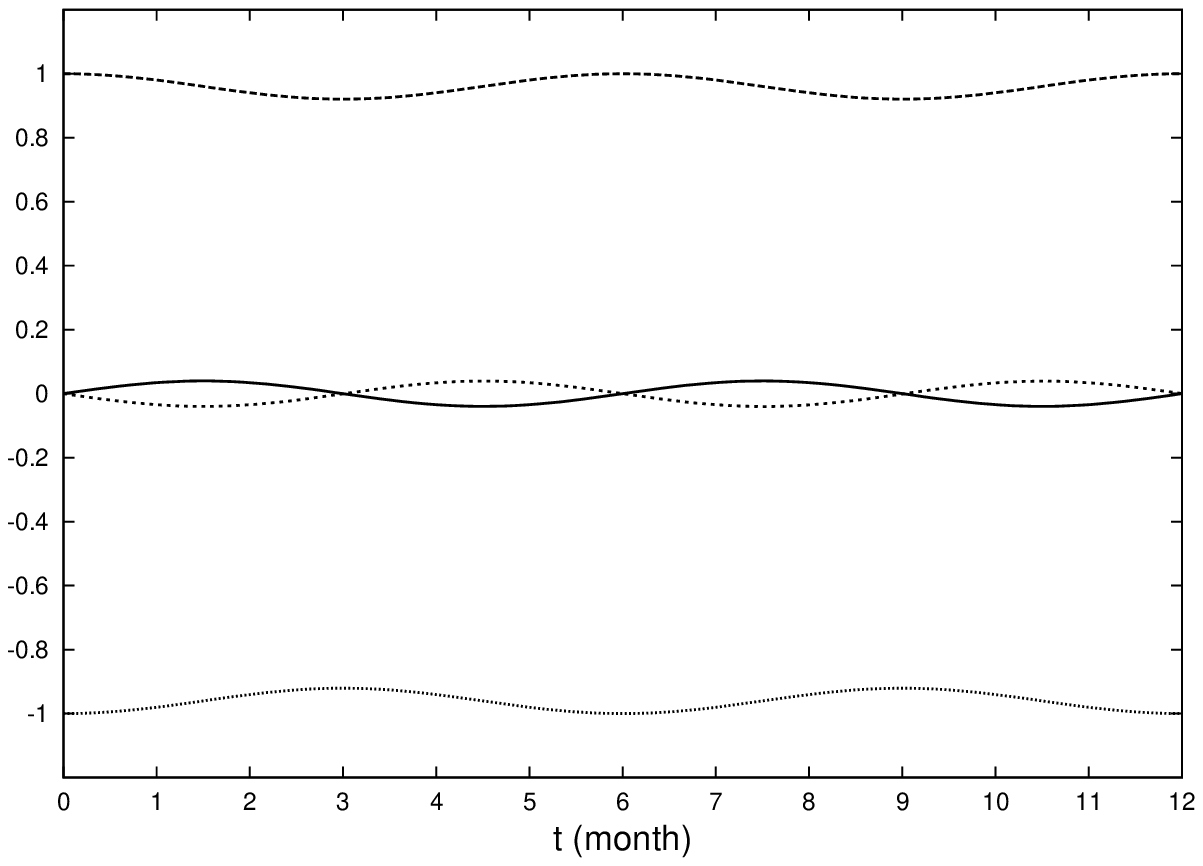}
\caption{ 
Seasonal variation in phase differences by CS effects 
on the same quantum interferometer (on a horizontal plane) 
as the solid curve in Fig. \ref{f2}. 
The winter solstice day is chosen as 0 month 
along the horizontal axis and the summer solstice 
corresponds to six months. 
Upper panel: full data points. 
Bottom panel: solid, long dashed, short dashed and dotted curves 
for 0, 6, 12 and 18 hours of each day, respectively. 
}
\label{f3}
\end{figure}

Figures \ref{f2} and \ref{f3} show 
numerical calculations of time variations in the phase difference 
for the Earth's parameters 
such as the inclination angle of the Earth's axis 
$I_E$, % \approx 23 \mbox{[deg]}$, 
the mean orbital angular velocity 
$\Omega_E$, % \approx 2\pi/365 [\mbox{rad}\cdot\mbox{day}^{-1}]$, 
the spin rate  
$\omega_E$, % \approx 2\pi/24 [\mbox{rad}\cdot\mbox{hr}^{-1}]$. 
where the eccentricity of the Earth orbit makes a tiny input. 
The magnitude of the variations in the phase shift 
by the inclination of the Earth's axis is expected to be 
$\sim O(\sin I_E) \sim O(0.1)$. 

Finally, we mention whether other quantum gravity effects 
could be present in the phase shift. 
Eq. (\ref{CS-action}) is the first Parity-violating term 
in a series of curvature corrections. 
There are probably other terms that would be cubic 
and higher order corrections. 
The next term would induce a correction 
$\sim \dot f^2 (\nabla \times V)^2$ 
in Eq. (\ref{g0i}), so that the part in the second parenthesis 
of Eq. (\ref{Delta-CS2}) could include $\dot f^2 c^{-2} 10^{-30}$ 
at the next order. 
Hence, the next (and higher order) terms can be safely neglected. 
Other couplings motivated by quantum gravity on the quantum
interference are left as a future work.

%\section{Conclusion}
\noindent \emph{Discussion and Conclusion.--- }
We considered effects of CS gravity 
on a quantum interferometer.  
The CS effect in the phase shift has an analogy in the AB one. 
The CS effects turn out to be 
dependent on the latitude and direction 
of the interferometer on the Earth 
in orbital motion around the Sun. 
The CS dependence is different from the general relativistic one. 
Daily and seasonal variations in phase shifts, 
independent of the wavelength, are thus 
suggested with an estimate of the size of the effects.  
Numerical studies 
for an interferometer at middle latitudes 
are left as a future work \cite{OYA}. 

Current measurements of phase shifts in neutron interferometry 
do not report any anomalous (daily nor seasonal) variations  
with 
phase measurement accuracy 
at $O(10^{-3})$. 
Current neutron interferometry, 
therefore, places a bound on CS gravity as 
$\dot f c^{-1} < 10^{0} \mbox{s}$ ($\dot f < 10^{5} \mbox{km}$), 
which is worse by three digits than the constraint 
$\dot f c^{-1} < 10^{-3} \mbox{s}$ 
by the classical experiment GPB (Gravity Probe B) 
\cite{AY1, GPB}
and also LAGEOS \cite{Smith}. 
Future progress in quantum technology 
may improve the bound. 
A bound comparable to the GPB limit would be placed, 
if neutron interferometry were sufficiently improved for 
$\Delta \times S^{-1}$ (nearly by three digits), 
for instance $\sim 5$ meters arm length 
and $\sim 10^{-4}$ phase measurement accuracy.  
It is awaited. 
Experimental setups usually suffer from many other seasonal variations. 
Lacking a signal, a constraint may be placed on $\dot f$. 
In the presence of a signal, on the other hand, 
one would have to eliminate all other possible sources of 
seasonal variability.

Finally, we mention briefly a possible path toward 
the desired technology improvement. 
Neutron interferometers are typically made from a single large crystal 
of silicon, 20 to 60 cm or more in length. 
Modern semiconductor technology allows large single-crystal silicon 
boules to be easily grown \cite{Book}. 
In the near future, therefore, neutron interferometry 
with a few-meter arm length might become available. 
For improving the phase measurement accuracy, technological challenges 
are in progress. 
For observations of the topological Aharonov-Casher effect, 
Werner and his collaborators have already obtained the result 
for the measured phase shift with a one-sigma statistical error bar 
as $\pm 0.34$ mrad $\sim O(10^{-4})$, where approximately, 
a total of 500 000 000 neutrons were 
counted in the interferograms over a period of 2 years 
(See \cite{Werner} for a review of observations of Aharonov-Bohm 
effects by neutron interferometry). 
Hence, CS gravity effects depending on the direction 
(such as the northeast) of the interferometer could be tested, 
if the Werner's interferometer had a longer arm (a few meters) 
and rotated. 
%For more reducing statistical error bars, the intensity of neutron 
%beams should be increased. 
Next, the CS dependence on the latitude might be tested, 
if another interferometer with the same capability 
were built in a different latitude (for instance on the equator). 

Furthermore, Seki and his collaborators have recently 
developed a multilayer cold-neutron interferometer 
and experimentally the phase measurement accuracy of $0.01$ rad, 
where they used only $1.5 \times 10^5$ neutrons in a short time 
$\sim 49$ hours \cite{Seki}. 
Such a technology might be used for a future experimental test 
of CS-type seasonal variations. 
However, it seems insufficient for experimental tests 
of daily variations. 
For a daily variation test, a breakthrough 
in neutron interferometry is needed. 
Motivated by quantum computations, for instance, 
Pushin and his collaborators 
have demonstrated experimentally how quantum-error-correcting codes 
may be used to improve experimental designs of quantum devices 
to achieve noise suppression in neutron interferometry \cite{Pushin}.

We would like to thank N. Yunes and S. Takeuchi  
for the useful discussions. 
% This work was supported in part (K.Y.) 
% by the JSPS fellowship. 


\begin{thebibliography}{99}
\bibitem{AY3}
References \cite{AY1, AY2} adopt the geometrical units 
($G=c=1$), with which quantum experimentalists 
may not be familiar. 
Hence, $G$ and $c$ are kept in this paper.
\end{thebibliography}
\end{document}